\documentclass[%
 reprint,
 amsmath,amssymb,
]{revtex4-1}

\usepackage{graphicx}
\usepackage{dcolumn}
\usepackage{bm}

\begin{document}


\title{
Electronic structure depiction of magnetic origin in  BaTiO$_{3-\delta}$  thin film:\\ A combined experimental and first-principles based investigation}

\author{
Supriyo Majumder$^{a}$,
\
Pooja Basera$^{b}$,
\
Malvika Tripathi$^{a}$,
\
R. J. Choudhary$^{a,*}$
\
Saswata Bhattacharya$^{b,\#}$
\
Komal Bapna$^{c}$,
and
D. M. Phase$^{a}$.
\\
$^{a}$UGC DAE Consortium for Scientific Research, Indore 452001, India\\
$^{b}$Department of Physics, Indian Institute of Technology Delhi, New Delhi 110016, India\\
$^{c}$National Physical Laboratory, New Delhi 110012, India
}
\address{Author(s) to whom correspondence should be addressed. Electronic mail(s): $^{*}$ram@csr.res.in, $^{\#}$saswata@physics.iitd.ac.in}






\begin{abstract}
With the motive of unraveling the origin of native vacancy induced magnetization in ferroelectric perovskite oxide systems, here we explore the consequences of electronic structure modification in magnetic ordering of oxygen deficient epitaxial BaTiO$_{3-\delta}$ thin films. Our adapted methodology employs state-of-the-art experimental approaches viz. photo-emission, photo-absorption spectroscopies, magnetometric measurements duly combined with first principles based theoretical methods within the frame work of density functional theory (DFT and DFT+\textit{U}) calculations. Oxygen vacancy (O$ _{V} $) is observed leading  partial population of Ti 3\textit{d} (t$_{2g}$), which induces defect state in electronic structure near the Fermi level and reduces the band gap. The oxygen deficient  BaTiO$_{2.75} $ film reveals Mott-Hubbard insulator characteristic, in contrast to  the band gap insulating nature of the stoichiometric BaTiO$ _{3}$. The observed magnetic ordering is attributed to the asymmetric distribution of spin polarized charge density in the vicinity of O$ _{V} $ site which originates unequal magnetic moment values at first and second nearest neighboring Ti sites, respectively. Hereby, we present an exclusive method for maneuvering the band gap and on-site electron correlation energy with consequences on magnetic properties of BaTiO$_{3-\delta}$ system, which can open a gateway for designing novel single phase multiferroic system.

\end{abstract}

\maketitle


\section{INTRODUCTION}{\label{sec:INTRODUCTION}
\paragraph*{}
In oxide systems, the physical and chemical properties are greatly affected by stoichiometry, lattice strain and vacancy induced defect concentrations \cite{Chen Zhou et al. 2013, Kennedy et al. 2014, Shimada et al. 2012, Chung et al. 2002, Zhu et al. 2007, Li Choudhury et al. 2006, Chen et al. 2013, Xing et al. 2008}. The fact that mutually exclusive ferroelectricity and ferromagnetism can co-exist in a perovskite ferroelectric system by introducing native vacancies, is of emerging relevance from both technological and fundamental aspects. Owing to the excellent ferroelectric properties, Pb(Zr,Ti)O$_{3} $ (PZT) thin films have gained tremendous attention till now. However, the toxicity of lead oxide is a major hindrance towards its usage in technological applications \cite{Shieh et al. 2007, Qiao et al. 2008}. On the other hand lead free BaTiO$_{3}$ (BTO) has emerged as an environment friendly substitute of PZT\cite{Shieh et al. 2007, Qiao et al. 2008, Liao et al. 2017, Tang et al. 2004, ZhouZ et al. 2013} and in addition, promising candidate for the technological demand as capacitors, photo-capacitors, energy storage capacitors, FERAM, optical modulator, electro-optical switch, wave guide \cite{Chen et al. 2013, Xing et al. 2008,  Qiao et al. 2008, Yano et al. 1994, Liu et al. 2018, Parizi et al. 2014, Sangsub Kim et al. 1996, Towner et al. 2003, Li et al. 2001}, piezoelectric nanogenerators \cite{Shin et al. 2017, SuoG et al. 2016, YanJ et al. 2016}, dye-sensitized solar cell electrodes (either in single or in composite form) \cite{OKAMOTO et al. 2014, OKAMOTO et al. 2015, Hatameh et al. 2017, ZhangL et al. 2008}, sensors \cite{ZhaoG et al. 2018} etc. 

\paragraph*{}
Various theoretical calculations have predicted that modulating defects generated by creating vacancies at crystal sites can induce magnetic ordering in BTO\cite{Raeliarijaona et al. 2017, D Cao et al. 2009, D Cao et al. 2011}. The magnitude of defect generated magnetization is noticed to be higher for Titanium vacancy in comparison to Oxygen vacancy (O$_{V} $) in BTO system \cite{Raeliarijaona et al. 2017, D Cao et al. 2009, D Cao et al. 2011}. However, O$_{V}$ has lower formation energy than Ti vacancy and is a better controllable parameter, offering a suitable pathway to study the induced magnetism in an otherwise non-magnetic BTO. O$_{V}$ can be considered as self doping process, which effectively acts as additional electron donor in the system, leading to a wide variety of functionalities \cite{Chen et al. 2013, Xing et al. 2008, Li et al. 2001, Eom et al. 2017}. Some recent studies demonstrate that nano-crystalline BTO can simultaneously exhibit ferroelectricity and ferromagnetism arising from core and O$ _{V} $ at surface of the nano-particles, respectively \cite{A. Sundaresan et al. 2009I, R. V. K. Mangalam et al. 2009I, F. Yang et al. 2010, D. Qin et al. 2010, M. Wang et al. 2010, S. G. Bahoosh et al. 2011, H. Liu et al. 2011, S Ramakanth et al. 2014}. Despite several research endeavors, the nature and mechanism of induced magnetic ordering is still not properly understood. As it is already transpired by the theoretical predictions that the modification of electronic structure due to O$_{V}$ holds the key to understand the induced magnetism, indeed, a systematic combined experimental and theoretical study is lacking to visualize the alteration in electronic structure caused by O$_{V}$ exclusively.  
\paragraph*{}
O$_{V}$  can be generated in a very controlled manner while fabricating thin films using pulsed laser deposition (PLD) technique by tuning the oxygen partial pressure (OPP) during the growth process. Here in this work, we elaborate how the O$_{V}$ in epitaxial BaTiO$ _{2.75} $ thin films significantly modifies the electronic structure across E$_F$, reduces the band gap, shows the Mott-Hubbard insulating character, and consequently leads toward the low temperature magnetic ordering, by means of X-ray photoelectron spectroscopy (XPS), valence band spectroscopy (VBS) and X-ray absorption near edge spectroscopy (XANES), magnetometric measurements along with first-principles based spin-resolved density functional theory (DFT and DFT+\textit{U}) calculations.

\section{METHODS}

\subsection{Experimental}

\paragraph{Thin film growth and structural characterization:}
Epitaxial BTO thin films (thickness $ \sim $ 150 nm $ \pm $ 5) are deposited on [001] oriented LaAlO$_{3} $ (LAO) single crystal substrate, by PLD using a KrF excimer laser system (Lambda Physik, wavelength 248 nm, pulse width 20 ns). The single phase stoichiometric bulk target of BTO for thin film growth, is synthesized by conventional solid state reaction method. BTO thin films are grown at different OPP and corresponding nomenclature of films are: BL100 (OPP = 100 mTorr), BL50 (OPP = 50 mTorr) and BL25 (OPP = 25 mTorr). After deposition these films are cooled under same OPP conditions, as used during growth process. Deposition temperature is kept at 750$ ^{\circ} $C and target to substrate distance is set to 4.5 cm. The laser fluence at the target surface is kept at 1.8 J/cm$ ^{2} $. The phase and structural characterization of the films are done by $ \theta-2\theta $ scans using Cu K$ \alpha $ ($ \lambda$=1.54 {\AA}) X-ray diffraction (Bruker D2 Phaser Desktop Diffractometer). Reciprocal space mapping (RSM) is carried out using high resolution X-ray diffractometer (Bruker D8 Discover HRXRD). 

\paragraph{Magnetometric measurements and electronic structure probe:}Magnetization measurements are performed using MPMS 7 Tesla SQUID-VSM (Quantum Design Inc., USA). The chemical valence state of the elements present in the sample are investigated by XPS experiments using Al K$ \alpha $ lab-source (1486.7 eV) with Omicron energy analyzer (EA-125). For occupied density of states (DOS), valence band spectra (VBS) is recorded at angle integrated photoemission spectroscopy (AIPES) beamline, using synchrotron radiation source (Indus-1, Beam Line 2, RRCAT, Indore, India). Before XPS and VBS measurements, the film surface is cleaned by sputtering using 0.5 keV Ar$^{+} $ ions inside the sample preparation chamber where ultra-high vacuum ($\sim$10$^{-10} $ Torr) is maintained. The charging effect corrections in XPS and Fermi level alignment in VBS, are done by measuring C 1\textit{s} core level and Au foil (which is in thermal and electrical contact with the sample) VBS, respectively. XPS and VBS spectra are fitted with combined Lorentzian-Gaussian function and Shirley background. Unoccupied DOS is probed by XANES measurements in total electron yield (TEY) mode at O K edge using synchrotron source at soft X-ray absorption spectroscopy (SXAS) beamline (Indus-2, Beam Line 1, RRCAT, Indore, India). All these spectra are recorded at 300 K. The estimated experimental resolutions for VBS and XANES measurements across the measured energy range are about 0.3 eV and 0.25 eV, respectively.

\subsection{Theoretical}
\paragraph*{}Density functional theory (DFT) calculation is performed using the projector augmented plane-wave (PAW) by employing Vienna ab initio Simulation Package (VASP) \cite{Kresse et al. 1996, Kresse et al. 1999}. A model structure of BaTiO$_3$ consisting of 20 atoms with periodic boundary condition is constructed by 4$\times$1$\times$1 replication of the tetragonal BaTiO$_{3} $ unit cell (space group \textit{P4mm}).  Several variants of DFT exchange and correlation functionals viz. Local Density Approximation (LDA)\cite{Perdew et al. 1992}, Generalized Gradient approximation (GGA) in the form proposed by Perdew, Burke and Ernzerhof (PBE) \cite{Perdew et al. 1996}, LDA+\textit{U}, GGA+\textit{U} (where \textit{U} is Hubbard parameter and it's value is carefully chosen from experimental inputs) are tested. Due to presence of transition metal in our system, we have employed the DFT+\textit{U} approach. The basic idea behind DFT+\textit{U} is to treat the strong on-site Coulomb interaction of localized electrons, which is not correctly described by local or semi-local functionals viz. LDA or GGA. It can be described quite accurately at the cost of almost similar computational efforts as with LDA/GGA by involving an additional Hubbard-like term. The on-site Coulomb interactions are particularly strong for localized \textit{d} and \textit{f} electrons, but can be also important for \textit{p} localized orbitals. The strength of the on-site interactions are usually described by parameters \textit{U} (on site Coulomb) and \textit{J} (on site exchange). These parameters \textit{U} and \textit{J} can be extracted from ab-initio calculations, but usually are obtained semi-empirically. For BaTiO$_3$ system, in order to describe precisely the strong on-site Coulomb repulsion among the localized Ti 3\textit{d} electrons, we adopt the LDA+\textit{U} and GGA+\textit{U} formalisms for the exchange correlation term. The equilibrium geometries, band structure, the total and partial density of states (DOS and PDOS) are systematically investigated and analyzed in comparison with experiments and previous calculations. The structure is fully relaxed (both atomic position and cell size) upto 0.001 eV/$\textrm{\AA}$ force tolerance using conjugate gradient minimization with 2$\times$2$\times$2 K-mesh. The computationally optimized structure details are given in supplementary material (SM). In our calculations, Gaussian smearing is used and the pressure convergence value is set to 0.0001 eV/\AA$ ^{3} $. For electronic structure energy calculations, the Brillouin Zone is sampled with a 8$\times$8$\times$8 Monkhorst-Pack \cite{Monkhorst et al. 1976} K-mesh with 0.01 meV energy tolerance. Note that spin-orbit coupling (SOC) is also included while performing the energy calculations. In all our calculations, the plane wave energy cut-off is set to 600 eV.

\section{RESULTS AND DISCUSSION}
\subsection{Structural properties}
\paragraph*{}
Figure \ref{XRD RSM}(a) shows the X-ray diffraction $\theta-2\theta$ scans of BTO films on LAO substrate, grown at different OPP values (100, 50 and 25 mTorr). These films are in single phase and [00l] oriented. From literature \cite{Li et al. 2001, W Li et al. 1999}, it is known that the BTO films grown by PLD at or above 150 mTorr of OPP, have a stoichiometry close to the ideal value. When these films are deposited at reduced oxygen environment, the films become oxygen deficient \cite{W Li et al. 1999}. As our prime focus is to investigate the electronic structure and magnetic properties of highest possible oxygen deficient BTO system, we have also deposited thin film at OPP of 5 mTorr. However, its XRD revealed signature of some secondary phase, which could not be indexed with BaTiO$ _{3} $. Therefore, that sample is not considered for further analysis. The calculated in-plane lattice mismatch with the substrate defined as ($a_{BTO}$-a$_{LAO}$)/a$_{LAO}$ is $ \sim $ + 5.36\%. Such lattice mismatch will cause in-plane compressive strain in the film during the initial growth stage. From the diffraction pattern, it can be noted that the [00l] peak shifts towards smaller $ 2 \theta $ values with decreasing OPP. This suggests the increment of the out-of-plane lattice parameter with the decrease of OPP during growth of BTO thin film. 

\begin{figure*}[t]
\centering
\includegraphics[angle=0,width=0.8\textwidth]{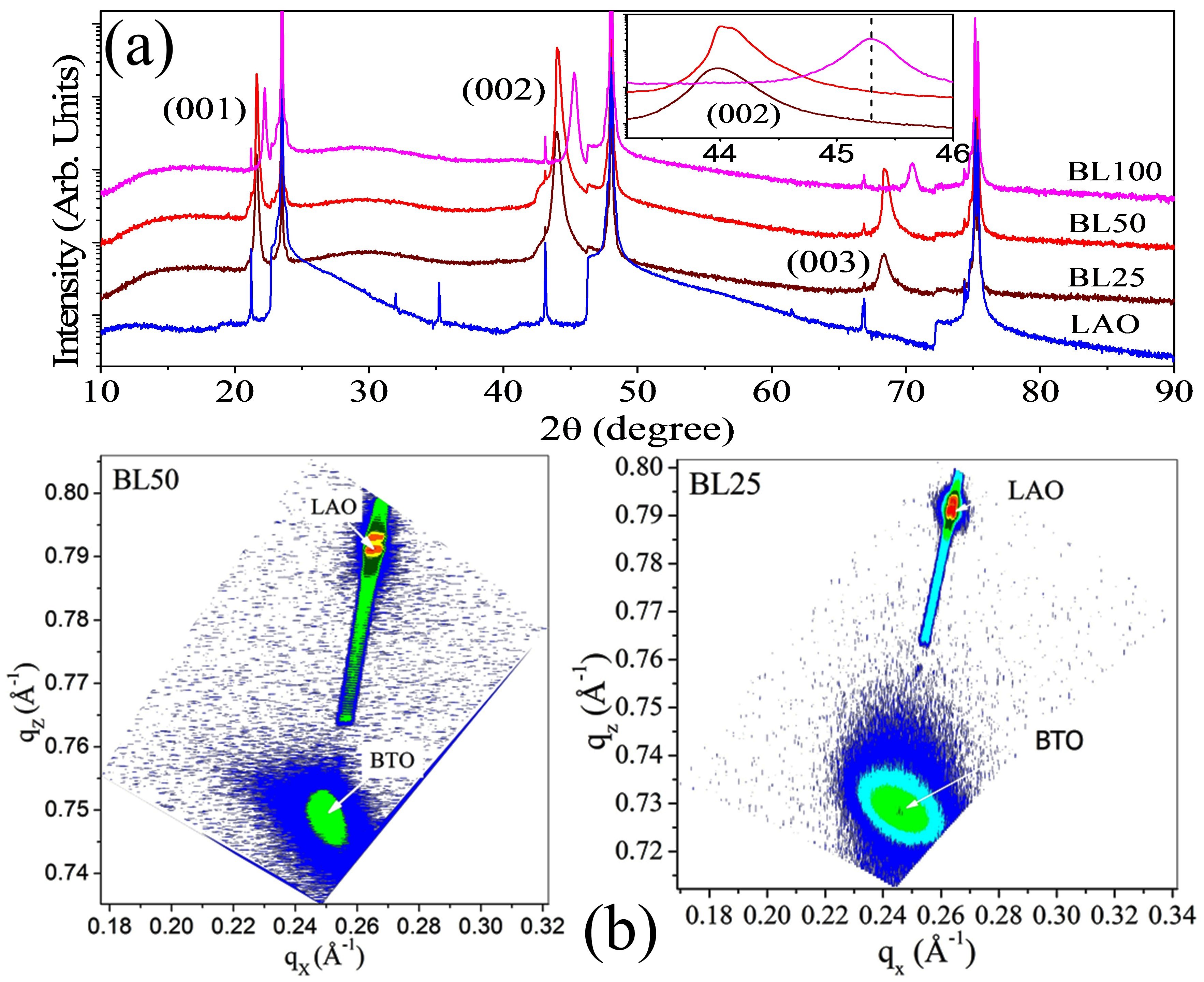}
\caption{(a): XRD ($ \theta-2\theta $) patterns of BTO thin films deposited at different oxygen partial pressures on (001) LAO substrate along with the (001) LAO substrate, inset displays the (002) peak of BTO films. (b): Reciprocal space maps around asymmetric (103) plane of BTO/LAO films. Arrows show the most intense coordinates in (q$ _{x} $, q$ _{z} $) domain correspond to BTO film and LAO substrate.}\label{XRD RSM}
\end{figure*}

\paragraph*{}
To obtain the nature of substrate induced strain present in the film, RSM around asymmetric (103) reflection is performed, as shown in Fig.\ref{XRD RSM}(b) for samples BL25 and BL50. It is clear from these RSM plots that both of the films are epitaxial in nature. Here the q$_{x} $ and q$_{z} $ axis correspond to the in-plane and out-of-plane directions, respectively. Let (q$_{xf} $, q$_{zf} $) and (q$_{xs} $, q$_{zs} $) denote the maximum intensity coordinates for the film and substrate, respectively. In both the cases (q$_{xf} $, q$_{zf} $) is sufficiently shifted from (q$_{xs} $, q$_{zs} $), indicating that the films are relaxed. The large lattice mismatch ($ \sim $ + 5.36\%) between bulk BTO and LAO substrate and a thickness higher than the critical one to sustain substrate induced strain, may be the possible reason for such strain relaxation. The in-plane lattice parameter `a' calculated from q$_{xf} $, for both the films are found close to the BTO bulk in-plane lattice parameter value. But, the q$_{zf}$ values of BL50 is significantly higher than the q$_{zf}$ value corresponding to BL25, suggesting an increase of out-of-plane lattice parameter `c' with decrease in OPP. In reduced oxygen environment O vacancies created in BTO films are compensated by lowering the oxidation state of some Ti ions from 4+ to 3+ (as shown later). The presence of Ti$^{3+}$ may be the possible reason for such elongation of out-of-plane lattice parameter (c), as Ti$^{3+}$ has larger ionic radius than Ti$^{4+}$.


\subsection{Electronic properties}

\subsubsection{Chemical valence state}

\paragraph*{}
For the electronic structure studies, we chose BL25 sample which was deposited in single phase under lowest OPP and hence should posses maximum O$ _{V} $ defect states among the grown films. To investigate the effect of oxygen deficiency in chemical valence state of the elements, we have performed XPS measurements of BL25 sample which is shown in Fig.\ref{Slide2}(a, b).
\begin{figure*}[t]
\centering
\includegraphics[angle=0,width=0.9\textwidth]{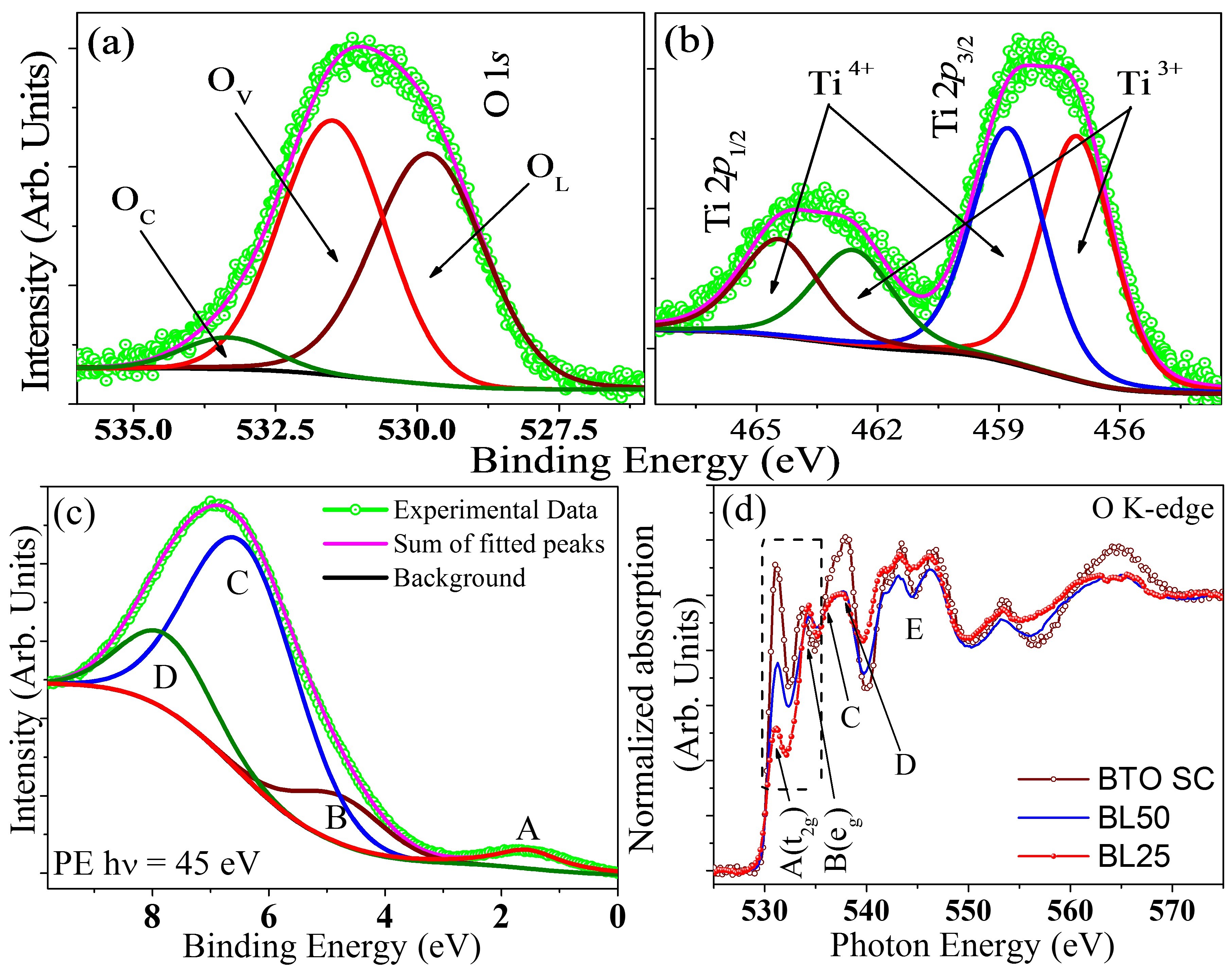}
\caption{XPS spectra of BL25 film, (a): O 1\textit{s}, indicating the presence of three types of oxygen, O$ _{L} $, O$ _{V} $ and O$ _{C} $ (see text); (b): Ti 2\textit{p}, indicating the presence of both Ti$ ^{4+} $ and Ti$ ^{3+} $ states. (c): Valence band spectra of BL25 recorded at photon energy of h$ \nu $=45 eV. (d): O K edge XANES spectra recorded at TEY mode for BTO single crystal, BL50 and BL25 samples.}\label{Slide2}
\end{figure*}
Deconvolution of the O 1\textit{s} XPS spectra in Fig.\ref{Slide2}(a) indicates the presence of three possible features having binding energies centered at 529.8, 531.5 and 533.3 eV, which are attributed to lattice oxygen (O$_{L} $) bound to Ti, O vacancies or defects (O$_{V} $) and the combined chemisorbed oxygen (O$_{C} $) species, respectively \cite{Hashimoto et al. 2015}. The core lines of Ti 2\textit{p} are splited into 2\textit{p$_{3/2}$} and 2\textit{p$_{1/2}$} due to spin orbit interaction, (Fig.\ref{Slide2}(b)) which can be deconvoluted by fitting the spectrum with peaks contribution from both Ti$^{3+} $ and Ti$^{4+} $ states, centered at binding energies 457.0 eV (Ti$^{3+} $ 2\textit{p}$ _{3/2} $), 462.6 eV (Ti$^{3+} $ 2\textit{p}$ _{1/2} $), 458.7 eV (Ti$ ^{4+} $ 2\textit{p}$ _{3/2} $) and 464.4 eV (Ti$^{4+} $ 2\textit{p}$ _{1/2} $), respectively \cite{Bapna et al. 2011}. From these XPS data it can be concluded that as expected, the BL25 film is highly oxygen deficient and these O vacancies are compensated by lowering the oxidation state of some Ti atoms, from 4+ to 3+. It is also noticed that the integrated intensity of Ti$^{3+}$ state is very close to Ti$^{4+}$ state and on average these XPS data reveal that the Ti$^{4+}$ : Ti$^{3+}$ content is in 1 : 1 ratio. Thus the effective chemical formula of BL25 sample will be; [ \{  Ba(Ti$ ^{4+}_{x} $Ti$ ^{3+}_{1-x} $)O$_{3-\delta} $ \}, such that x $ \approx $ (1-x) and x+2$ \delta $-1 $ \approx $ 0 ]  $ \cong $ Ba(Ti$ ^{4+}_{0.5} $Ti$ ^{3+}_{0.5} $)O$_{2.75} $ which will yield, BaTi$ ^{3.5+} $O$ _{2.75} $ with $ \delta $ = 0.25. This stoichiometry is maintained across the thickness as confirmed from the depth dependent XPS measurements (not shown here).


\subsubsection{Valence band spectroscopy}

\paragraph*{}
To analyze the occupied density of states near Fermi level, valence band spectroscopy measurements were performed. Figure \ref{Slide2}(c) shows the fitted VBS of BL25 recorded at incident photon energy of 45 eV. Deconvolution of this VB spectra reveals the four features A, B, C and D centered at binding energies 1.5, 4.7, 6.4 and 7.7 eV, respectively \cite{Hudson et al. 1993}. The feature B is assigned as O 2\textit{p} - O 2\textit{p} non-bonding states, while features C and D are ascribed to Ti 3\textit{d} - O 2\textit{p} bonding $ dp\sigma $ and $ dp\pi $ states, respectively \cite{S. W. Robey et al. 1996}. Feature A has been attributed to Ti$^{3+} $ 3\textit{d} defect state \cite{Bapna et al. 2011, S. W. Robey et al. 1996, Bapna APL et al. 2011}. It should be noted that feature A is not observed in the VBS of stochiometric BaTiO$ _{3} $ \cite{Hudson et al. 1993}. Therefore, the observance of feature A can be assigned to the presence of oxygen vacancies. O vacancy causes partial change in band filling from $ d^{0} $ (in stoichiometric BaTiO$_{3} $: Ti$^{4+} $) to $ d^{1} $ (partially populated 3\textit{d} in BaTiO$_{3-\delta} $: mixed Ti$^{3+} $ \& Ti$^{4+} $), which causes the appearance of  new feature A near Fermi level.

\subsubsection{X-ray absorption near edge spectroscopy}

\paragraph*{}
To probe the unoccupied electronic density of states of these oxygen deficient films, we have performed X-ray absorption near edge spectra measurements at the O K edge as shown in Fig.\ref{Slide2}(d), which gives information about the hybridization of transition metal and oxygen states \cite{Groot et al. 1989}. For comparison we have also measured XANES of BTO single crystal. The features marked by A, B and C in Fig.\ref{Slide2}(d) correspond to the hybridized states between O 2\textit{p} and Ti 3\textit{d}, while feature D is assigned to O 2\textit{p} derived states hybridized with Ba 5\textit{d} \cite{Mastelaro et al. 2015}. Feature E can be identified as excitation from Ti 4\textit{sp} - O 2\textit{p} \cite{Suntivich et al. 2014}. The two peaks labelled as A and B can be identified as t$_{2g} $ and e$_{g} $ bands, respectively \cite{Groot et al. 1989} arising due to crystal field splitting.
\paragraph*{}
It is observed that the intensity ratio of t$_{2g} $ related feature (A) with e$_{g} $ feature (B) varies as, I(t$_{2g} $)/I(e$_{g} $):BTO single crystal $ > $ I(t$_{2g} $)/I(e$_{g} $):BL50 $ > $ I(t$_{2g} $)/I(e$_{g} $):BL25. In general the intensity ratio of t$_{2g} $ and e$_{g} $ can be considered as a rough estimate of \textit{d} electron count \cite{Groot et al. 1989}. O vacancy induced doped electrons start filling empty t$_{2g} $ (Ti 3\textit{d}) orbitals and as a result the number of unoccupied states reduces. This is further confirmed by reduction in I(t$_{2g} $)/I(e$_{g} $) intensity ratio with increasing O vacancies. The intensity ratio I(t$_{2g} $)/I(e$_{g} $) is also related to hybridization strength \cite{Groot et al. 1989}. The drop of intensity ratio I(t$_{2g} $)/I(e$_{g} $) indicates that the hybridization between Ti 3\textit{d} and O 2\textit{p} decreases with increasing O vacancies. It should be noted that O vacancy induces a reduction in Ti valence state. Lowering the oxidation states leads to the increase in charge transfer energy $ \Delta $ \cite{Bocquet et al. 1992}. Thus the higher oxygen deficient films with higher Ti$^{3+} $ content should show weaker hybridization between Ti 3\textit{d} and O 2\textit{p}, since the hybridization strength is inversely proportional to $ \Delta $ \cite{Suntivich et al. 2014}, which is reflected in the XANES result. 

\begin{figure*}[t]
\centering
\includegraphics[angle=0,width=0.6\textwidth]{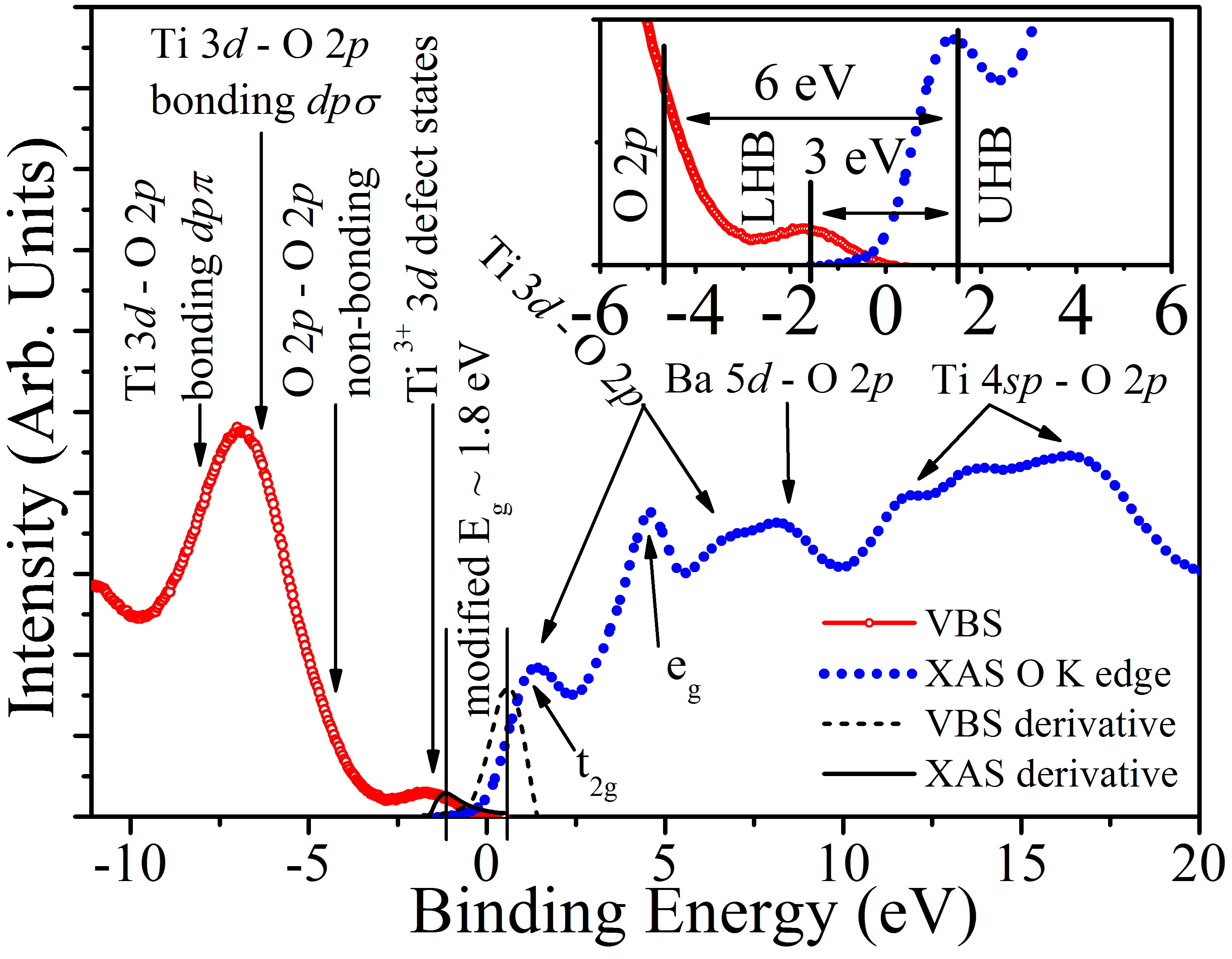}
\caption{Schematic electronic band structure of BL25 film. The presence of band gap defect state modifies the band gap and electronic structure of the system. Inset shows the estimated \textit{U} and $ \Delta $ (see text).}\label{Experimental electronic structure BL25}
\end{figure*}

\subsubsection{Electronic structure near Fermi level}

\paragraph*{}
Finally, using XPS O 1\textit{s} core level binding energy, the VBS and XANES spectra are brought to common energy scale \cite{Galakhov et al. 2010}, to get a schematic of electronic band structure (shown in Fig.\ref{Experimental electronic structure BL25}) of BL25 sample. It is known that the stochiometric BaTiO$ _{3} $ is a band-gap (3.2 eV) insulator \cite{Kolodiazhnyi et al. 2010}. However, in BaTiO$ _{3-\delta} $ thin film, as evident from Fig.\ref{Experimental electrornic structure BL25}, due to O vacancies, the overall electronic structure is greatly modified across the Fermi level. The presence of Ti$ ^{3+} $ 3\textit{d} defect state in band structure near the Fermi level causes a reduction in the band gap (from 3.2 eV to $ \sim $ 1.8 eV). The estimated band width of this metal 3\textit{d} defect feature is about \textit{W} = 1.4 eV. From this electronic structure, the Ti$ ^{3+} $ 3\textit{d} defect state in the valence band and the Ti 3\textit{d} t$ _{2g} $ feature in conduction band can be identified as the spectroscopic signature of lower Hubbard band (LHB) and upper Hubbard band (UHB), respectively. From the energy separation between the UHB and LHB , we have estimated the on-site \textit{d-d} Coulomb interaction energy \textit{U} to be approximately equal to 3 eV\cite{Maiti et al. 2000-II}. Similarly, the ligand to metal charge transfer energy $ \Delta $, can be estimated as the energy difference in UHB and O 2\textit{p} band \cite{Maiti et al. 2000-II} and it takes a value of about 6 eV in our system. Here we have observed \textit{U}/\textit{W} $ > $ 1 and \textit{U} $< \Delta $, thus the oxygen deficient BaTiO$_{3-\delta} $ system should be categorized as a Mott-Hubbard type insulator. The resistivity behavior (SM, Fig.S3) confirms that the increase in O$ _{V} $ concentration increases the conductivity, even-though the system remains insulating in nature.
 


\subsection{Magnetic properties}

\paragraph*{}
Now to study the effect of O vacancy on the magnetic properties of BaTiO$_{3-\delta} $ system, we have performed dc magnetization measurements. Figure \ref{MT MH}(a) shows the M(T) curves in field-cooled warming (FCW) mode in presence of  $\mu_0$H =100 Oe applied magnetic field in a direction parallel to the surface of the film. In order to estimate the ordering temperature, here we have determined inflection temperature point in temperature derivative of magnetization, which is an indication of magnetic transition \cite{Weiwei Li et al. 2013, H. Z. Guo et al. 2008}. As shown in the inset of Fig.\ref{MT MH}(a), the derivative of magnetization with respect to temperature ($ \frac{dM}{dT} $) has the minima at T $ \approx $ 43.9 K for BL25 and 42.3 K for BL50. Therefore, we have considered the magnetic transition temperatures as 43.9 K and 42.3 K for BL25 and BL50, respectively. The change in the O$ _{V} $ concentration significantly affects the magnetic moment values but the ordering temperatures do not change as such. The magnetic transition temperature does not show any significant field dependency (SM, Fig.S1), but as expected, the magnetic moment value increases with increasing the measuring magnetic field. It should be noted that no anomalous magnetic ordering is observed when we repeated the same measurement with stoichiometric single crystal (SC) of BTO. With decreasing OPP the O$_{V} $ and O$_{V} $ induced Ti$^{3+} $ concentrations in the films increase. As the magnetic moment is contributed from these O$_{V} $ induced Ti$^{3+} $, the BL25 thin film shows moment value one order higher than BL50. To ascertain the hysteric nature of these BaTiO$_{3-\delta} $ thin films, M(H) isotherms were collected at different temperatures, as shown in Fig.\ref{MT MH}(b). The diamagnetic contribution from the substrate is generated from extrapolating the high filed linear part of M(H) curves and subtracted with the overall magnetic moment. Finite value of coercivity (Hc $ \sim $ 200 Oe) is observed for both of the films. As the minor impurities in substrates may also cause some unexpected magnetic ordering, we have also performed magnetization measurements of bare LAO substrate (SM, Fig.S2) treated under identical thermal and vacuum conditions which were used for film (BL25) fabrication. The magnetization behaviors of the substrate show typical dia-magnetic nature, confirming the observed magnetic results of BTO/LAO samples are intrinsic magnetic properties of BaTiO$ _{3-\delta} $ system. As expected, BL25 (M$_{S} \sim $ 0.57$ \mu $B/f.u.) shows much higher saturation magnetization than BL50 (M$_{S} \sim $ 0.27$ \mu $B/f.u.).

\begin{figure*}[t]
\centering
\includegraphics[angle=0,width=0.9\textwidth]{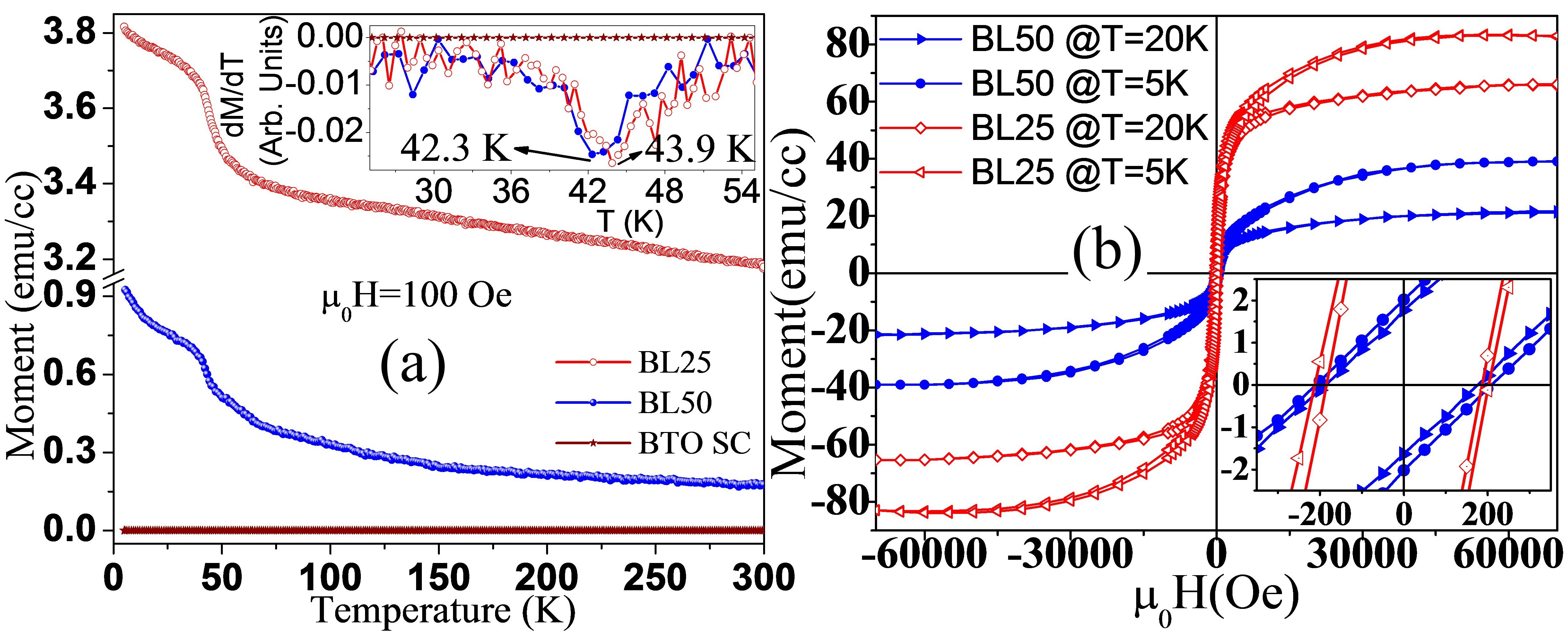}
\caption{(a): FCW cycle of M(T) curves at applied magnetic field 100 Oe. Inset shows magnetization derivative as a function of tempareture. (b): M-H isotherms measured at 20 K and 5 K temperatures.}\label{MT MH}
\end{figure*}

\paragraph*{}
As discussed earlier, there are lot of debates regarding the localization of two free electrons generated by the formation of one O vacancy. Cao \textit{et al}.\cite{D Cao et al. 2009}, suggested that O vacancy induced magnetic moment in bulk BaTiO$_{3} $ arrises from unpaired electrons at neighboring of Ti $ t_{2g} $ states. According to first principles calculations based on GGA formalism in ref. Cao \textit{et al}. 2009\cite{D Cao et al. 2009}, O vacancy mediated magnetic moments at nearest and second nearest neighbors of Ti atoms are about 0.22-0.28 $ \mu $B and 0.06-0.09 $ \mu $B, respectively, which give rise a total magnetic moment of about 1.27-1.54 $ \mu $B in tetragonal phase of BaTiO$_{3} $. In another theoretical report Cao \textit{et al}.\cite{D Cao et al. 2011} also studied the effect of O vacancies on magnetic properties of (001) cubic BaTiO$_{3} $ thin film using DFT calculations and claimed that the O vacancy in Ba-O terminated surface can cause magnetism which is due to the spin-polarization of electrons localized at vacancy basin. The GGA+\textit{U} calculations in ref. Cao \textit{et al}. 2011\cite{D Cao et al. 2011} showed that the presence of O vacancies can induce magnetic moment about 1.12 $ \mu $B (with \textit{U} = 0) and 1.93 $ \mu $B (with \textit{U} = 3 eV). Whereas, Raeliarijaona \textit{et al}.\cite{Raeliarijaona et al. 2017}, showed that O vacancy induced magnetization in bulk BTO have its origin in the spin polarization of the itinerant electrons at Ti 3\textit{d} (t$_{2g} $) orbitals. The LDA+\textit{U} calculations for different \textit{U} in ref. Raeliarijaona \textit{et al}. 2017\cite{Raeliarijaona et al. 2017}, yield the magnetic moment value 0.25, 0.5 and 1.5 $ \mu $B for \textit{U} = 0, 3 and 6, respectively. All these reports claim that ferromagnetism is energetically more favorable than antiferromagnetism in O deficient BaTiO$_{3-\delta} $ system. 

\paragraph*{}
From the electronic structure it is evident that O vacancy leads to modification in VB as well as variation in feature related Ti 3\textit{d} t$ _{2g} $ state in the conduction band. This clearly underlines that O vacancy results in occupancy of 3\textit{d} states of neighboring Ti atoms converting them from Ti$ ^{4+} $ to Ti$ ^{3+} $. In the crystal structure of BTO, an O atom is shared between two neighboring Ti atoms. If both these Ti atoms accept an electron each due to O vacancy, they will be converted into Ti$ ^{3+} $. It is shown in previous studies that Ti$^{3+}$O$^{2-}$Ti$^{3+}$ exchange interactions display antiferromagnetic ordering \cite{Tokura et al. 1993, Kumagai et al. 1993}. However, in the present study the observed magnetic properties preclude such possibility of colinear antiferromagnetism. Thus it is quite reasonable to assume that in the present case most of the electrons available due to O vacancies are localized at the adjacent Ti site, transforming them from Ti$ ^{4+} $ to Ti$ ^{3+} $, whereas some of them are also distributed at next Ti atoms. This will create an asymmetric charge distribution at Ti sites in proximity of O vacancy, viz. Ti$ ^{4-\epsilon} $-O-Ti$ ^{3+\epsilon} $-O$ _{V} $-Ti$ ^{3+\epsilon} $-O-Ti$ ^{4-\epsilon} $ chains. The exchange interaction between these local moments will lead to magnetic ordering with uncompensated magnetization. O vacancy in BTO induced n-type carrier doping (as discussed later). Such defect carriers mediate the magnetic interactions by forming bound magnetic polaron (BMP) \cite{Yuan Hua Lin et al. 2009, Coey et al. 2005} with local magnetic moment due to Ti$ ^{3+} $. Depending on the carrier concentration and magnetic impurity concentration, such BMPs overlap and lead to long range magnetic ordering between Ti$ ^{4-\epsilon} $ and Ti$ ^{3+\epsilon} $ ions. As the O vacancy concentration increases, more number of bound magnetic polarons are formed yielding a larger value of magnetic moment with increasing O vacancy.

\begin{figure*}[t]
\centering
\includegraphics[angle=0,width=0.9\textwidth]{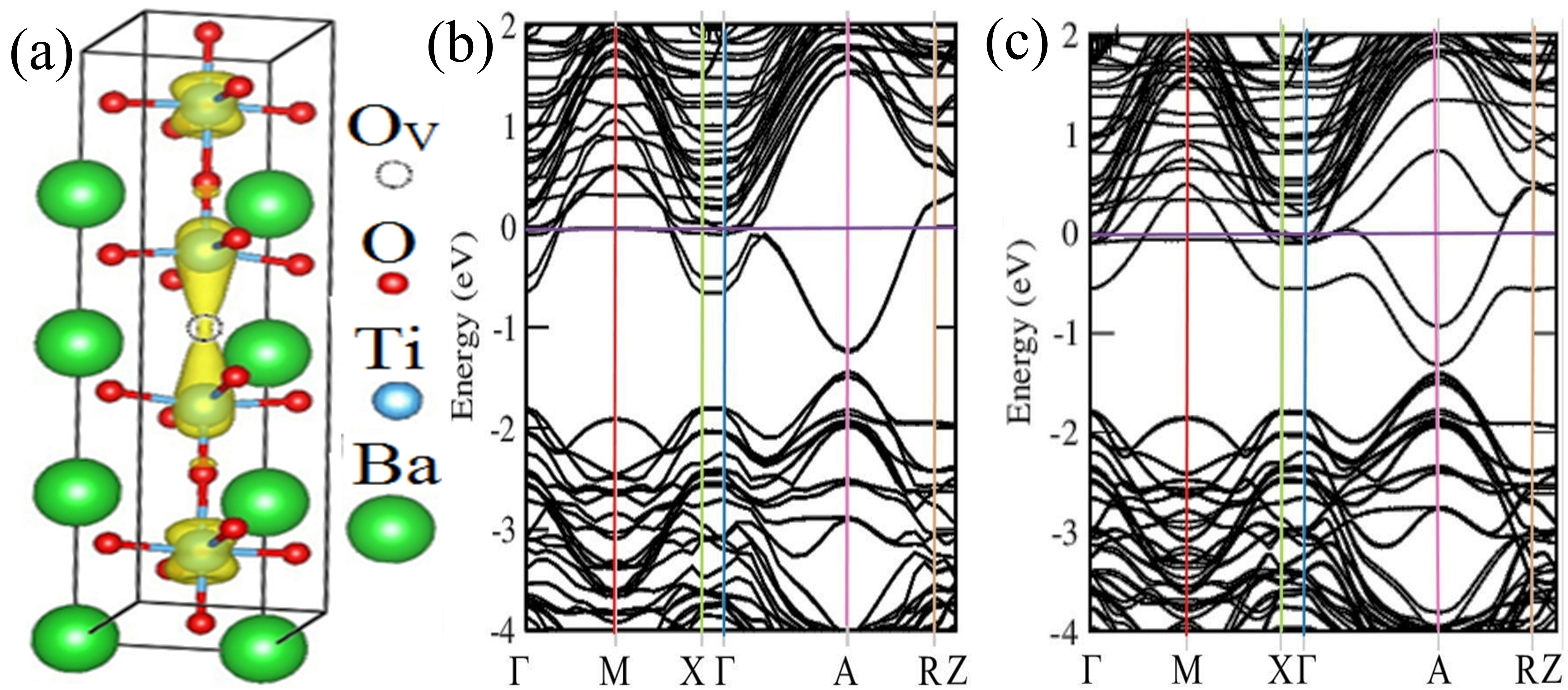}
\caption{(a): Spatial distribution of spin polarized charge density around oxygen vacancy in BaTiO$_{2.75}$ supercell. Electronic band structures of BaTiO$ _{2.75} $ composition simulated using; (b): LDA+\textit{U}, (c): GGA+\textit{U} functionals. }\label{Supercell band structure}
\end{figure*}




\subsection{First principles calculations}

\paragraph*{}
The DFT calculation is carried out on BaTiO$_{3} $ (stiochiometric) and BaTiO$_{2.75} $ (oxygen deficient) system. For simulation of BaTiO$_{2.75} $ composition, one oxygen atom is removed from the center of the supercell (Fig.\ref{Supercell band structure}(a)). To get a better insight we have used DFT+\textit{U} approach with \textit{U} = 3 eV, which was estimated from present spectroscopic observations (shown in Fig.\ref{Experimental electronic structure BL25}). This on-site Coulomb interaction \textit{U} is added to Ti site. Theoretically calculated band structure of BaTiO$_{2.75} $ system using LDA+\textit{U} and GGA+\textit{U} formalism are shown in Fig.\ref{Supercell band structure}(b, c). For band structure we have plotted the Kohn-Sham energy eigenvalues calculated along the high symmetry path i.e., $ \Gamma $ (0 0 0), M (0.5 0.5 0), X (0.5 0 0), $ \Gamma $ (0 0 0), A (0.5 0.5 0.5), R (0.5 0 0.5), Z (0 0 0.5). The DFT band gap of the stiochiometric BaTiO$_{3} $ (tetragonal \textit{P4mm}) is $ \sim $ 1.78 eV from LDA and $ \sim $ 2.8 eV from GGA approaches (DFT underestimates the band gap), which is in agreement with previous calculations \cite{Padilla et al. 1997}. 
It is observed that O vacancy induces modification in the band structure and reduces the band gap of the system as compared to the stoichiometric case. Fig.\ref{Slide5}(a) (LDA, LDA+\textit{U} scheme) and Fig.\ref{Slide5}(b) (GGA, GGA+\textit{U} scheme) depict the simulated total and partial density of states (PDOS) for the stoichiometric and oxygen deficient BTO system. Comparing the stoichiometric and oxygen deficient case, one can identify some noticeable features, as following: (i) the presence of defect states in PDOS of Ti 3\textit{d} (very near to the Fermi energy) changes the overall DOS of the system and the orbital-resolved DOS of Ti (not shown here) confirms that these states consist of Ti 3\textit{d} t$ _{2g} $ orbitals, (ii) O vacancy causes a reduction in band gap of the system, and (iii) O vacancy leads to n-type doping in the system. These findings are well consistent with our experimental outcomes.

\begin{figure*}[t]
\centering
\includegraphics[angle=0,width=0.9\textwidth]{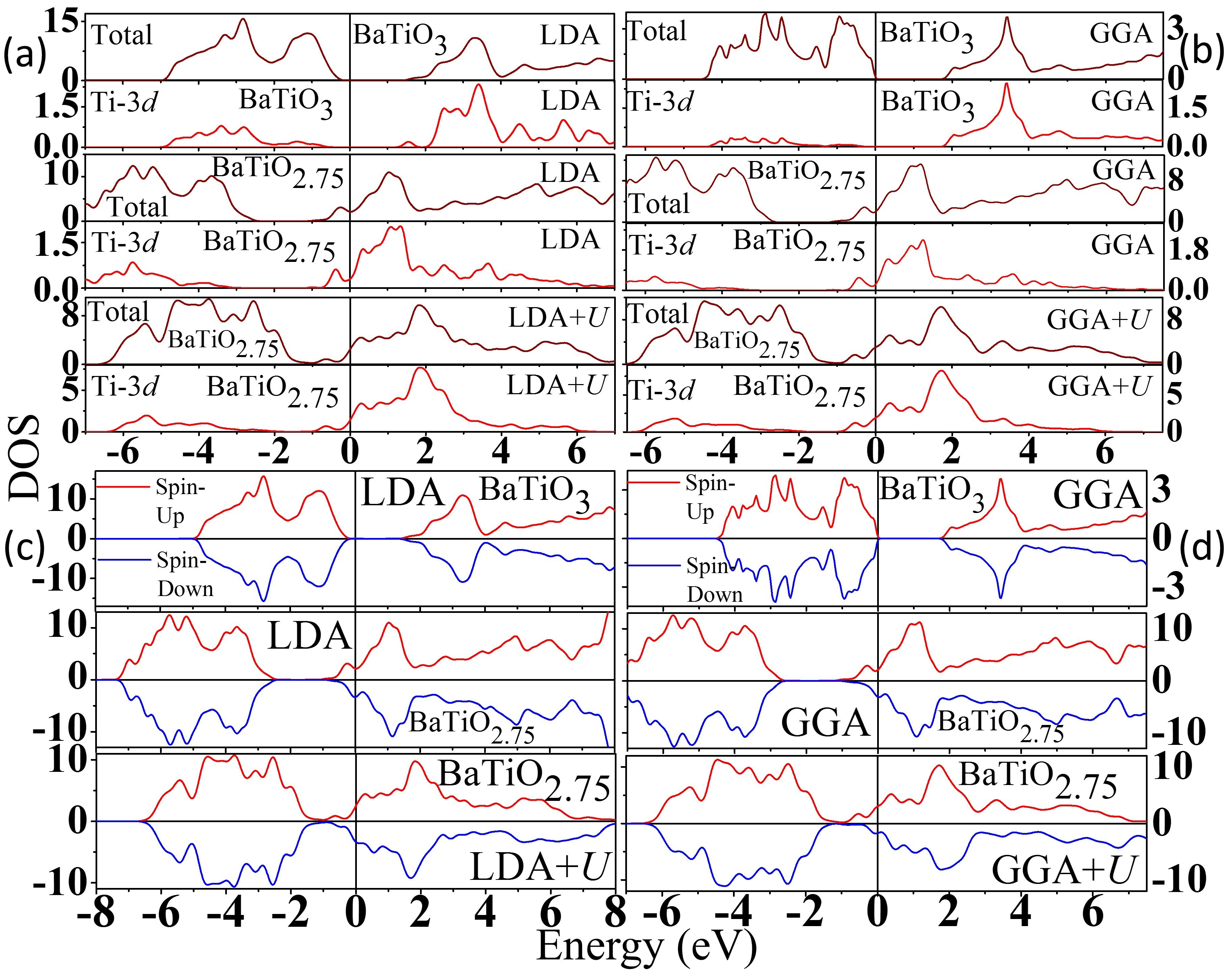}
\caption{Total and partial (Ti-3\textit{d}) DOS near the Fermi level for BaTiO$ _{3} $ and BaTiO$ _{2.75} $ compositions calculated using; (a): LDA and LDA+\textit{U} functionals, (b): GGA and GGA+\textit{U} functionals. Spin-resolved DOS near the Fermi level simulated using; (c): LDA and LDA+\textit{U} functionals, (d): GGA and GGA+\textit{U} functionals. }\label{Slide5}
\end{figure*}




\paragraph*{}
To understand the origin of magnetism, we have performed spin-resolved DFT calculations on stoichiometric BaTiO$_{3} $ and oxygen vacant BaTiO$_{2.75} $ compositions. The spin-resolved density of states near the Fermi energy level are plotted in Fig.\ref{Slide5}(c) and Fig.\ref{Slide5}(d) for different theoretical schemes LDA, LDA+\textit{U}, GGA and GGA+\textit{U}. Here, the spin-up and spin-down components are shown in positive and negative scale, respectively. Both of our LDA and GGA calculations show identical spin-up and spin-down DOS component for stoichiometric BaTiO$_{3} $ system, resulting a vanishing magnetic moment value. By focusing on the spin-resolved DOS obtained from LDA, LDA+\textit{U}, GGA and GGA+\textit{U} formalisms for  BaTiO$_{2.75} $ composition, one can clearly identify that here spin-up and spin-down DOS components are not symmetric i.e. they are spin-polarized. This spin polarization induces magnetism in oxygen deficient BaTiO$_{2.75} $ system. Inclusion of on-site Coulomb interaction \textit{U} in spin resolved DFT+\textit{U} calculations, results in more asymmetry in spin-up and spin-down DOS i.e. DFT+\textit{U} calculations yield higher magnetization. 

\paragraph*{}
To investigate the issue of charge localization induced by oxygen vacancy we have studied the spatial distribution of spin polarized charge density around the oxygen vacancy site (shown in Fig.\ref{Supercell band structure}(a)) in BaTiO$_{2.75} $ composition. Figure \ref{Supercell band structure}(a) shows that the spin polarized charge density is mostly spread at first nearest neighbor (FNN) Ti sites adjacent to the oxygen vacancy. A small part of the spin polarized charge density is also distributed at second nearest neighbor (SNN) Ti atoms. From GGA+\textit{U} approach we have found that magnetic moment in BaTiO$_{2.75} $ is originating from FNN and SNN Ti atoms neighboring to the oxygen vacancy and the calculated values of magnetic moments at FNN and SNN Ti sites are about, 0.44$ \mu_{B} $ and 0.12$ \mu_{B} $, respectively. These results further confirm our prediction of asymmetric charge distribution at Ti sites in the proximity of oxygen vacancy, which leads to asymmetric distribution of magnetic moment values. After creating oxygen vacancy, the lattice was  allowed  to relax around the vacancy site. We have found that Ti atoms adjacent to the oxygen vacancy cite goes away from each other due to coulomb repulsion and as a result of this the supercell gets elongated (shown in Fig.\ref{Supercell band structure}(a)). Oxygen vacancy induced charge imbalance is compensated by lowering the valence state of adjacent Ti atoms. The lower valency Ti ions having larger ionic radius will push away each other by coulomb repulsion. This is also corroborated by the elongated lattice parameter as observed from XRD and RSM analysis (shown in Fig.\ref{XRD RSM}). 

\paragraph*{}
From the aforementioned discussions, it is transpired that the oxygen vacancy greatly modifies the electronic structure of BTO. These modifications in electronic structure have huge bearings on the magnetic properties. It is observed that oxygen vacancy leads to (i) n-type doping in the system, (ii) partial occupancy of Ti 3\textit{d} t$ _{2g} $ states, (iii) reduction in band gap, (iv) change in insulating nature from band gap to Mott-Hubbard with on-site \textit{d-d} Coulomb repulsion energy \textit{U} being 3 eV, (v) induce magnetic ordering, (vi) formation of spin-polarized states near the Fermi energy level, (vii) inclusion of \textit{U} in DFT+\textit{U} calculations causes more spin-polarization in spin resolved DOS components and (viii) asymmetric distribution of spin-polarized charge densities at FNN and SNN Ti sites adjacent to the O$ _{V} $.  \\ 

\section{CONCLUSION}
\paragraph*{}
In summary, we have investigated the oxygen vacancy (O$ _{V} $) induced modifications in electronic structure on BaTiO$ _{3-\delta} $ thin film system, using combined experimental and theoretical tools to interpret the origin of magnetic ordering. O$ _{V} $ related charge imbalances are compensated by lowering the oxidation state of some Ti atoms from 4+ to 3+ states. Both of our experimental and theoretical investigation confirm partial population of Ti 3\textit{d} t$ _{2g} $ state in BaTiO$ _{3-\delta} $ system, which modifies the electronic structure across the Fermi level resulting a reduction in band gap of the system. By estimating \textit{d-d} Coulomb interaction energy \textit{U} and ligand to metal charge transfer energy $ \Delta $, we have categorized the oxygen deficient BaTiO$ _{2.75} $ as a Mott-Hubbard insulator. Bulk magneto-metric study reveals that oxygen vacancy induces anomalous magnetic ordering in BaTiO$ _{3-\delta} $ thin film systems. On the basis of observed signatures of the asymmetric charge distribution in the vicinity of O$ _{V} $ site, we have explained the intriguing magnetic ordering in BaTiO$ _{3-\delta} $ thin films accounting the interaction of Ti$ ^{4-\epsilon} $-O-Ti$ ^{3+\epsilon} $-O$ _{V} $-Ti$ ^{3+\epsilon} $-O-Ti$ ^{4-\epsilon} $ chains. The spin resolved density of states and spatial distribution of spin polarized charge density in the proximity of the vacancy site confirm our interpretation regarding the induced magnetism. The present work will in general help to tune the oxygen vacancy induced magnetic ordering in prototype systems.
\\
\section*{ACKNOWLEDGMENTS}
Authors acknowledge Dr. V. Raghavendra Reddy for RSM; Mr. Tarachand Patel and Dr. Gunadhor Okram for resistivity, Mr. Avinash Wadikar, Mr. Sharad Karwal and Mr. Rakesh Sah for helping in measurements using synchrotron beamline at Indus - I \& II, RRCAT, India. SM would like to thank Mr. Gyanendra Panchal and Mr. Anupam Jana for providing necessary discussions. PB acknowledges UGC, India, for the junior research fellowship [grant no. 20/12/2015(ii)EU- V]. PB and SB acknowledge the High Performance Computing (HPC) facility at IIT Delhi for computational resources. SB acknowledges the financial support from YSS-SERB research grant, DST, India (grant no. YSS/2015/001209).

\end{document}